\documentclass[letterpaper]{article} % DO NOT CHANGE THIS
\usepackage{aaai2026}  % DO NOT CHANGE THIS
\usepackage{times}  % DO NOT CHANGE THIS
\usepackage{helvet}  % DO NOT CHANGE THIS
\usepackage{courier}  % DO NOT CHANGE THIS
\usepackage[hyphens]{url}  % DO NOT CHANGE THIS
\usepackage{graphicx} % DO NOT CHANGE THIS
\usepackage{amssymb}
\usepackage{natbib}  % DO NOT CHANGE THIS AND DO NOT ADD ANY OPTIONS TO IT
\usepackage{caption} % DO NOT CHANGE THIS AND DO NOT ADD ANY OPTIONS TO IT
\usepackage[table]{xcolor}
\usepackage{colortbl} % Optional, for additional color options

\usepackage{algorithm}
\usepackage{algorithmic}

\usepackage{newfloat}
\usepackage{listings}
\DeclareCaptionStyle{ruled}{labelfont=normalfont,labelsep=colon,strut=off} % DO NOT CHANGE THIS
\floatstyle{ruled}
\newfloat{listing}{tb}{lst}{}
\floatname{listing}{Listing}
\title{Learning to Use AI for Learning: Teaching Responsible Use of AI Chatbot to K-12 Students Through an AI Literacy Module}
\author {
    Ruiwei Xiao\textsuperscript{\rm 1},
    Xinying Hou\textsuperscript{\rm 2},
    Ying-Jui Tseng\textsuperscript{\rm 1},
    Hsuan Nieu\textsuperscript{\rm 3},
    Guanze Liao\textsuperscript{\rm 3},
    John Stamper\textsuperscript{\rm 1},
    Kenneth R. Koedinger\textsuperscript{\rm 1}
}
\affiliations {
    \textsuperscript{\rm 1}Carnegie Mellon University\\
    \textsuperscript{\rm 2}University of Michigan\\
    \textsuperscript{\rm 3}National Tsing Hua University\\
    ruiweix@cs.cmu.edu
}

\begin{document}

\maketitle

\begin{abstract}
As Artificial Intelligence (AI) becomes increasingly integrated into daily life, there is a growing need to equip the next generation with the ability to apply, interact with, evaluate, and collaborate with AI systems responsibly. Prior research highlights the urgent demand from K-12 educators to teach students the ethical and effective use of AI for learning. To address this need, we designed a Large-Language Model (LLM)-based module to teach prompting literacy. This includes scenario-based deliberate practice activities with direct interaction with intelligent LLM agents, aiming to foster secondary school students' responsible engagement with AI chatbots. We conducted two iterations of classroom deployment in 11 authentic secondary education classrooms, and evaluated 1) AI-based auto-grader's capability; 2) students' prompting performance and confidence changes towards using AI for learning; and 3) the quality of learning and assessment materials. Results indicated that the AI-based auto-grader could grade student-written prompts with satisfactory quality. In addition, the instructional materials supported students in improving their prompting skills through practice and led to positive shifts in their perceptions of using AI for learning. Furthermore, data from Study 1 informed assessment revisions in Study 2. Analyses of item difficulty and discrimination in Study 2 showed that True/False and open-ended questions could measure prompting literacy more effectively than multiple-choice questions for our target learners. These promising outcomes highlight the potential for broader deployment and highlight the need for broader studies to assess learning effectiveness and assessment design.
\end{abstract}

% The contributions of this study include: 1) \textbf{Advancing pedagogical methods}: through analysis of the results, we summarized design implications to foster the development of AI Literacy education; 2) \textbf{Technical contributions}: examining the capabilities of LLMs in assessment and feedback generation in prompt literacy education context; 3) \textbf{AI Literacy learning and assessment materials}: we enriched AI Literacy learning resources by building an AI-enhanced online learning platform and open-sourcing our learning and assessment materials.  


\section{Introduction}
The increasing presence of AI in everyday life has amplified the need for effective techniques for human–AI interaction, including within educational settings \cite{van2019popstar}. As AI products have increased adaptability and ease of use, students can access them easily in different contexts \cite{hou2025exploring}, including at home and school \cite{yim2024artificial}. Some AI commercial products, like ChatGPT, have been introduced to K-12 educational settings, aiming to personalize student learning experience \cite{zhang2024systematic, nayak2023teaching}. Apart from its potential values, there are growing concerns about utilizing generative AI chatbot technologies including increasing cheating \cite{lee2024cheating, li2025unseen, xiao2025teachers} and the low quality of information retrieval in learning settings \cite{wang2023unleashing, kazemitabaar2023novices}.

While AI chatbots hold promise for K-12 education, few instructional interventions, to our knowledge, focus on teaching prompting literacy to K-12 students. Prompts are natural language instruction inputs that are used to communicate with AI chatbot technologies \cite{lo2023clear}. Prompting literacy includes helping students understand \textit{\textbf{what}} AI can do to support learning, \textit{\textbf{when}} to use AI effectively, and \textit{\textbf{how}} to craft prompts for different types of assistance. While the potential of LLMs is promising, their current effectiveness remains heavily dependent on the quality of the prompts they receive \cite{ekin2023prompt}. By designing and engineering them carefully, prompts can play a key role in shaping the AI's responses and guiding the output of LLMs to provide desired results \cite{lo2023clear, white2023prompt}. As prompts become an important end-user communication channel with AI chatbots, scholars in various educational domains argue that the ability to create effective prompts that generate desired information is now an essential skill for students \cite{denny2023conversing, korzynski2023artificial, mesko2023prompt, woo2024exploring}. Therefore, equipping students with prompting literacy is crucial to help students effectively communicate with AI chatbots during learning.

This work presents a web-based interactive instructional system to improve secondary-education (middle- and high-school) students' prompting literacy. It applies active learning \cite{bonwell1991active} and experiential learning \cite{ng2023ai} methods to engage students with prompting practice in three hypothetical learning scenarios. After a student writes a prompt, an AI auto-grader evaluates key dimensions and delivers immediate, detailed feedback. We hypothesized that practicing prompt writing in this way could help students learn to create effective prompts, supporting learning and confidence in AI use. We first outlined the module design. Then, we reported our classroom evaluation results. Specific research questions were listed under each evaluation study. Finally, we proposed ways to improve prompting literacy instruction.

\section{Related Work}
\subsection{Prompting literacy in K-12 classrooms}
It is believed that the ability to understand and use AI (or the lack thereof) will fuel the next digital divide in education \cite{trucano2023ai}. The emergence of generative AI makes such technologies more accessible to the general public and underscores that it is more urgent than ever to regulate the responsible and effective use of such technologies \cite{williamson2024time}. While K-12 teachers expressed their timely needs for AI competency materials for their students, especially prompting instructions \cite{lozano2023education}, the majority of current K-12 AI Literacy implementations emphasize understanding AI technologies than responsible use of AI. Additionally, existing AI competency frameworks primarily target adults in the workforce \cite{mesko2023prompt,tseng2024assessing, zamfirescu2023johnny} and higher-education \cite{denny2023conversing, giray2023prompt}. As a focused area within AI literacy, prompting literacy has gained increasing attention since 2023 \cite{hwang2023prompt, lo2023clear, maloy2024prompt, dennison2024consumers, xiao2025improving}. However, the lack of emphasis on K-12 learners persists, as few studies specifically target secondary education students, despite growing concerns about student AI use in K–12 education \cite{li2025unseen, uanachain2025generative}.

\subsection{Learn-by-Doing with Elaborated Immediate Feedback}

This study’s instructional design is grounded in two foundational learning sciences principles: learning-by-doing \cite{schank2013learning} and elaborated immediate feedback \cite{wang2019elaborated}. Learning-by-doing emphasizes active engagement through direct interaction with tasks. Prior research has shown that this approach is more effective than passive methods (e.g., reading, video-watching) in online learning environments \cite{koedinger2015learning}. While learning-by-doing has been applied in prompting literacy education (e.g., Promptly \cite{denny2023promptly}), such efforts often lack elaborated immediate feedback, which may hinder effective learning or even lead to drop-out \cite{vasilyeva2008immediate}.

Recent advances in large language models (LLMs) offer new opportunities to address this challenge. LLMs have demonstrated strong performance in rubric-based grading across domains \cite{henkel2024can, nguyen2023evaluating} and are capable of generating personalized feedback for open-ended responses \cite{xiao2024exploring, nguyen2023evaluating}. For example, GPT-4 achieves up to 85\% grading accuracy on short-answer questions using zero-shot prompting strategies \cite{henkel2024can}. Building on these capabilities, our work integrates an AI auto-grading pipeline with prompt-based practice to close the gap in prompting literacy instruction—enabling learning-by-doing accompanied by elaborated, immediate feedback to support more efficient and effective learning.

\section{Prompting Literacy Module Overview}

\subsection{Learning Objectives} 
\begin{figure*}[ht]
  \centering
  \includegraphics[width=\linewidth]{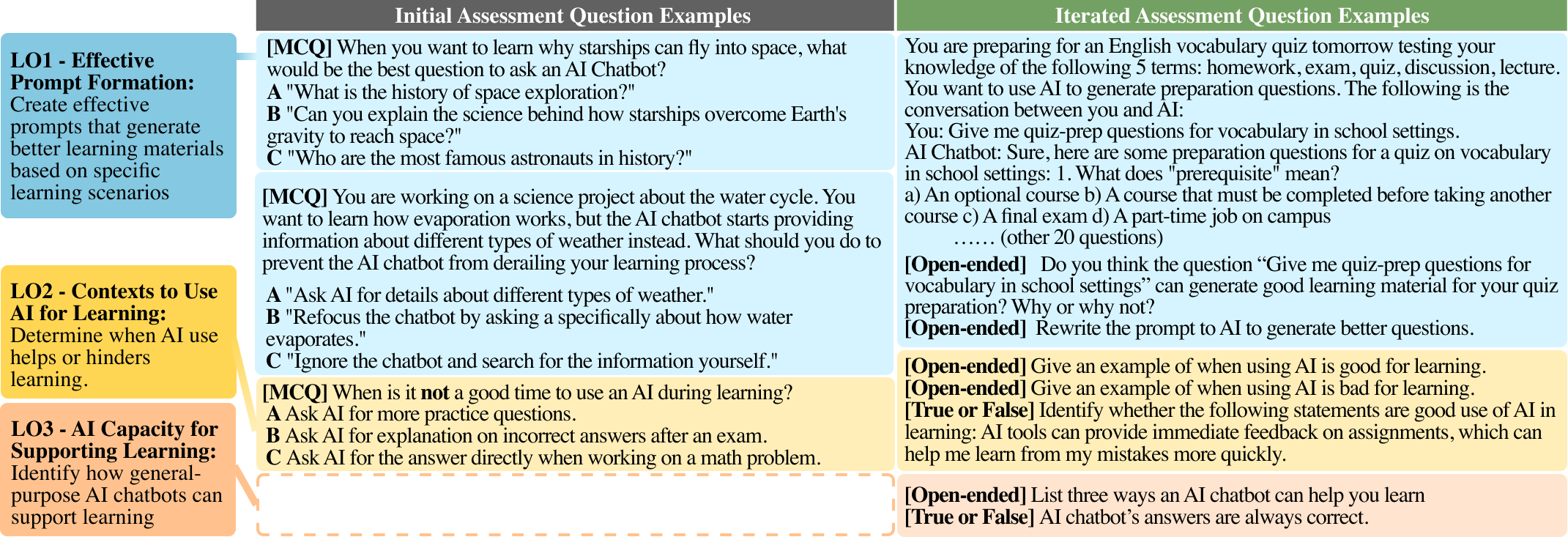}
  \caption{Initial and Iterated Assessment Questions with High-Level Learning Objectives (LO)}
  \label{fig:assessment_design}
\end{figure*}

Based on the real needs of teachers \cite{lee2024cheating} for instructions to teach students how to use AI for learning and the AI4K12 guideline \cite{touretzky2019special}, three learning objectives were derived: AI Capacity for Supporting Learning, Contexts to Use AI for Learning, and Effective Prompt Formation (Figure \ref{fig:assessment_design}).

\subsection{Learning Materials} 
We have designed and developed a web-based platform to support this prompting literacy practice. Students will first review the module’s learning objectives, then complete two conceptual practice questions with feedback and hints to explore how AI can support their learning. After these, students will get into the main practice activities.

\begin{figure*}[ht]
  \centering
  \includegraphics[width=0.95\linewidth]{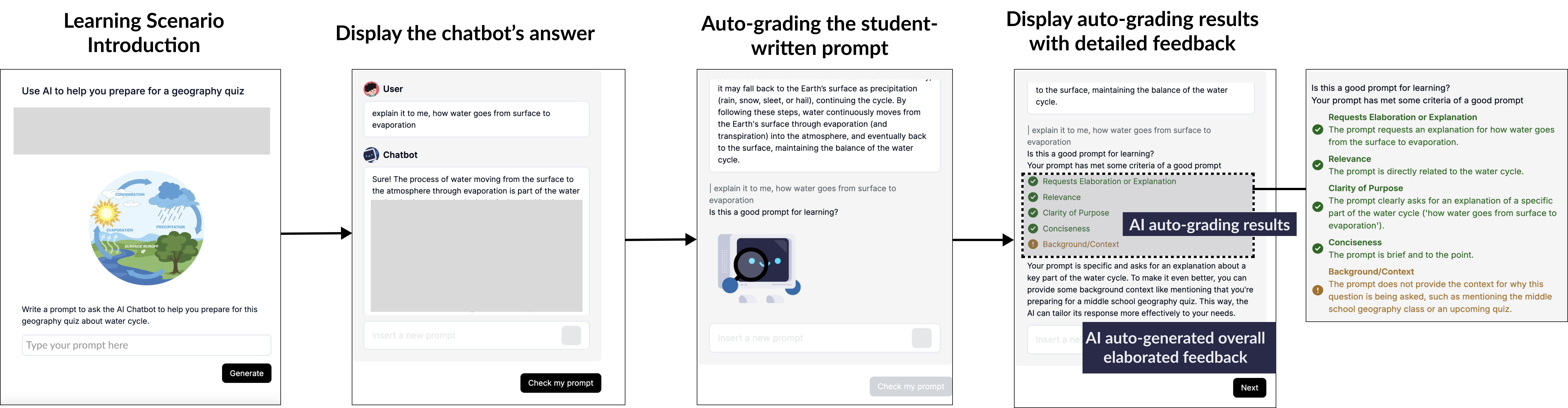}
  \caption{Students' practice pipeline in this interactive module}
  \label{flow}
\end{figure*}

The main practice is conducted using three hypothetical learning scenarios. Each practice has four steps (Figure \ref{flow}): 1) Introduce the detailed scenario to students, including the domain, the context, the student's current knowledge level, the learning goal the student is trying to achieve, and the resources the student has. 2) The student creates a prompt and clicks “generate" to receive answers from an AI chatbot. This mimics the authentic interaction with a general-purpose AI chatbot. 3) The student receives an answer from the AI chatbot based on the written prompt. 4) They can then click “Check my prompt" to receive an auto-graded result and a detailed explanation. This feedback is provided based on the preset evaluation dimensions for each scenario (Table \ref{tab: evaluation}). After receiving feedback, students can go back to craft a new prompt for the same practice or move on to the next practice. The detailed practice pipeline is shown in Figure \ref{flow}.

The three scenarios were designed under three secondary school subjects: biology, geography, and math. Each domain was paired with a unique instructional activity, such as \textit{“broaden knowledge beyond course requirements"}, \textit{“prepare for a quiz tomorrow"}, and \textit{“struggle with an assignment due tomorrow"}. Here is an example scenario:

\begin{quote}
\small
\textbf{Scenario 1: Use AI to extend your learning on biology} \textit{Yesterday, you learned 'what is a cell' in the biology class. You roughly understood what cells are and the main components in a cell. Now, as you are interested in biology, you want to broaden your understanding of cells. This isn't about course requirements, just a topic (cell) you're interested in learning more about. The only resource you have is your biology textbook. Write a prompt to ask the AI Chatbot to help you extend your knowledge about cells!}
\end{quote}

\subsection{Automatic Grading Dimensions} 
In our study, the requirements for a high-quality prompt vary depending on the instructional activity the student is pursuing (Table \ref{tab: evaluation}). For example, a good prompt in the \textit{homework struggle} situation should not explicitly ask for a direct answer. However, this criterion does not apply to other scenarios like \textit{course content extension}. Therefore, when designing the criteria for a high-quality prompt, each practice for each scenario has specific dimensions to meet. Also, each dimension has its own in-context descriptions rooted in the general definitions. In this study, we created our own grading dimensions based on the CLEAR framework for prompt creation \cite{lo2023clear} and other existing prompting recommendations \cite{mesko2023prompt}. Based on this grading rubric, for each student-written prompt, AI automatic grading needed to provide both a True/False categorical pass situation and a detailed explanation for this decision for each included dimension in that practice task. In addition, it was required to provide an detailed explanation for each dimension.

\begin{table*}[]
\fontsize{8.5pt}{9pt}\selectfont
\centering
\renewcommand{\arraystretch}{1.2} 
\rowcolors{2}{white}{gray!20}
\begin{tabular}{lllll}
\hline & \textbf{General definition}  & \textbf{Scenario 1} & \textbf{Scenario 2} & \textbf{Scenario 3} \\ \hline
\textbf{Relevance}   & The prompt is related to the topic  & \checkmark & \checkmark  & \checkmark  \\
\textbf{Clarity of Purpose}  & \begin{tabular}[c]{@{}l@{}}The prompt identifies a specific and clear purpose.\end{tabular}  & \checkmark & \checkmark & \checkmark \\
\textbf{Conciseness}  & The prompt itself is brief and concise.  & \checkmark     & \checkmark  & \checkmark \\
\textbf{Background/Context }      & \begin{tabular}[c]{@{}l@{}}The prompt explains why this question was asked, such as the background\end{tabular}   & \checkmark & \checkmark & \checkmark \\
\begin{tabular}[c]{@{}l@{}}\textbf{Request Elaboration}\\  \textbf{or Explanation}\end{tabular}      & \begin{tabular}[c]{@{}l@{}}The prompt requests elaboration, extension, or explanation  except for \\ seeking a direct response.\end{tabular} &  & \checkmark & \checkmark \\
\begin{tabular}[c]{@{}l@{}}\textbf{Not Explicitly Seeking} \\ \textbf{a Direct Response}\end{tabular} & \begin{tabular}[c]{@{}l@{}}The prompt does not ask for a solution to solve a problem or the answers \\  (what are x and y).\end{tabular}     &   & & \checkmark                                                                                            \\ \hline
\end{tabular}
\caption{Grading dimensions and associated tasks for student-written prompts. “\checkmark" means the LLM assesses if the student's prompt meets the definition with a binary response (Yes or No) and then gives corresponding grading explanations.}
\label{tab: evaluation}
\end{table*}

\section{Study 1}

As a novel activity type, we are interested in understanding students' perceptions of this activity, including what they learned from this activity, the interesting parts and the challenges encountered. In addition, as we integrated AI-based automatic grading into this procedure, we are interested in how it performed when grading students' written prompts using the preset evaluation dimensions. The first classroom evaluation study was conducted in June 2024 in 6 secondary-level classrooms in East Asia. We deployed this as an in-class practice with the local IRB (Institutional Review Board) approval. This resulted in valid data from 111 students. We explored the three RQs below:

\begin{itemize}
    \item \textbf{RQ1:} What is the accuracy of AI-based automatic grading for students’ written prompts?
    \item \textbf{RQ2:} How does this activity influence students’ prompting ability and confidence in using AI for future learning?
    \item \textbf{RQ3:} How do students perceive the learning experience, including its benefits and challenges?
\end{itemize}

\subsection{Study Procedure}
The practice was conducted at Information Technology classes with a teacher to help facilitate the whole process. After a researcher explained the practice purpose to students, they first conducted a pre-practice survey about their prior experience with AI. Then the students completed six multiple-choice questions (pre-test) on prompting literacy and suitable/unsuitable learning scenarios with AI. After completing the pre-practice survey and tests, the student moved to practice their prompting literacy in the system. Teachers were instructed to inform students that the auto-grader may make mistakes. 

After the practice, the student re-answered a 5-point Likert-scale survey question about \textit{“I know how to use AI to help me learn"} and the six multiple-choice questions again as the post-test. At the end, the student answered 6 open-ended survey questions to reflect on the learning experiences. The LLM used was OpenAI GPT-4o.

\subsection{Assessment Question Design}
Scenario-based assessments have been applied to measuring skills in varied domains, such as writing \cite{zhang2019scenario}, mathematics \cite{cayton2012technology} and science \cite{bergner2019process}. Such situated assessment experience can engage learners and facilitate deep thinking by presenting realistic, contextual problems that promote critical thinking and knowledge application \cite{clark2010formative, herrington2000instructional}. Furthermore, as our instructional goals were competency-based, scenario-based assessments are suitable to test students' abilities. Multiple Choice Questions (MCQs) enable automatic grading while providing consistent evaluations across diverse learner groups. When carefully crafted, MCQs can target any cognitive level in Bloom’s Taxonomy \cite{amer2006reflections}, making them a versatile and effective component of our assessment approach. Therefore, the assessment consisted of six MCQs. Each question presents three answer choices, with only one being correct. In Study 1, we chose MCQs as the only assessment question type and situated them into different scenarios based on the learning objectives.

\subsection{Data Collection and Analysis}
In \textit{Study 1}, we collected both students' written prompts and corresponding AI-grading results. To answer RQ1, we extracted students' last written prompt and graded them manually to assess AI auto-grading. After receiving 483 unique student-written prompts, two researchers used the same rubric to grade the students' written prompts (Table \ref{tab: evaluation}). They first randomly selected 15\% of the prompt data and conducted human grading based on the rubric. After that, the two researchers met to address the conflicts and iteratively refined the detailed grading book. Then they randomly selected another 15\% of the prompt data and both two researchers graded it again; the inter-rater reliability for this round reached higher than 0.92 (almost perfect agreement following \cite{mchugh2012interrater}). Then the two researchers met again to resolve the conflicts and graded the rest. We used human grading results for RQ2. For RQ3, we conducted a thematic analysis on student answers. One hundred and thirty-one students finished answering the self-reflection questions. Two researchers conducted a coding book iteratively to analyze their answers and reported the main themes.

\begin{table*}[]
\fontsize{8.5pt}{10pt}\selectfont
\centering
\begin{tabular}{lcccccccc}
\hline
 & \textbf{Relevance} & \textbf{Purpose} & \textbf{Conciseness} & \textbf{Background} & \textbf{Elaboration} & \textbf{No Answer} & \textbf{Overall} \\ 
\hline

Accuracy of Pass/Fail Classification (1 / 0) & 0.98 & 0.85 & 0.93 & 0.96 & 0.90 & 0.88 & -\\
Explanation Accuracy (1 / 0.5 / 0)  & 0.98 & 0.87 & 0.96 & 0.95 & 0.72 & 0.91 & 0.95\\

\hline
\end{tabular}
\caption{Grading Accuracy of the Auto-Grader} 
\label{tab: IRR}
\end{table*}

\subsection{RQ1: What is the accuracy of AI-based grading for students’ written prompts?}

\subsubsection{The grading system is able to grade and provide grading rationale with good quality in most categories.} Using the human labels as ground truth, in general, the AI auto-grader achieved an average of 0.92 accuracy when assessing students' prompts and generated high-quality, detailed feedback. The evaluation result is in Table \ref{tab: IRR}.

The auto-grader achieved a high accuracy (higher than 0.9) in \textit{Relevance}, \textit{Background}, and \textit{Elaboration} dimensions. For the inaccurate cases in these dimensions, one pattern is that the auto-grader tended to weigh heavily on some keywords but ignored other keywords \cite{nguyen2023evaluating}. For instance, one prompt was \textit{“You are preparing for a geography exam and have just learned the basic concepts of the water cycle Answer in Sanskrit."} The auto-grader gave it a False in \textit{Relevance} and explained as \textit{“The core of the problem lies in language choice and has nothing to do with the specific content of the water cycle."}.

The accuracy for \textit{Conciseness} and \textit{No Direct Answer} was 0.93 and 0.88. The conflicted cases for \textit{Conciseness} mostly included two types. First, the auto-grader focused on spelling errors and therefore graded them as False, while human graders were more tolerant about this. Second, prompts containing background information might be seen as non-concise by the auto-grader. For example, one student-written prompt was \textit{“Can you tell me in detail about your extracurricular knowledge about cells besides the definition of cells?"}. The auto-grader graded \textit{Conciseness} as False and explained it as \textit{“The question is relatively long and the sentences are not concise enough. It could be more refined."}

As for \textit{No Direct Answer}, sometimes the student asked a general question about \textit{“How to solve a two-variable linear equation?"}, but the auto-grader treated this as the student tried to ask for a direct answer \textit{indirectly} and labeled it as False. On the other hand, sometimes the student asked for \textit{“list the steps to solve this equation 10x+4y=3,-2x+10y=4"}, but auto-grader thought this prompt  \textit{“did not directly ask about the values of x and y, but rather asks for a solution."} and gave it True in \textit{No Direct Answer} instead.

Auto-grader received the lowest accuracy in \textit{Purpose} dimension (0.85). The main reason for false positive cases in this dimension is that the auto-grader might over-generate the keywords in a student-written prompt, and consider that as providing a clear purpose. For instance, it counted \textit{“I am a junior high school student. Please help me solve this problem using junior high school knowledge."} as True in \textit{Purpose} dimension and explained it as \textit{“the student's goal is to solve this problem."} However, simply asking for “solving this problem" without specifying the type or domain does not align with our rubric (\textit{“specific and clear purpose, e.g. explain the math concepts involved"}).

For true negative cases in \textit{Purpose}, sometimes the auto-grader mixed the criteria of \textit{Purpose} and \textit{No Answer}, and therefore considered those involving asking for an answer as False. Another reason is that auto-grader sometimes requires too detailed purpose, which is outside of our targeted scope. For example, one explanation to an AI-graded False case (\textit{“I want to know about DNA"}) is \textit{“Although DNA is mentioned in this question, the purpose is not clear. Students should be more specific about what they want to learn about DNA."} However, as the scenario is novice learners who only finished the first biology class about cells, the auto-grader should not expect the prompt purpose to be that detailed.

\subsection{RQ2: How does this activity influence students’ prompting ability and confidence in using AI for future learning?}

Students answered four 5-scale Likert questions as a pre-survey before doing the pre-test. After finishing the learning activities, they answered the last question in the pre-survey again. Specific questions and the student response distribution are shown in Figure \ref{fig:merged}.

\subsubsection{Students improved at embedding background in prompts with practice.}
Given that students' scores for each prompt dimension are dichotomous categorical data (Yes-1/No-0) and we wanted to understand how students improved from Q1 to Q3, we conducted a series of McNemar tests by comparing the common four prompting dimensions across all three practice questions (\textit{Relevance, Conciseness, Background, and Purpose}). We found no significant differences in \textit{Relevance} dimension (\textit{p} $>$ .90), \textit{Conciseness} dimension (\textit{p} = .286), and \textit{Purpose} dimension (\textit{p} = .617). Students generally performed well in these dimensions even in the first question. However, most of the students lacked the awareness of embedding background and context information in their prompt at the first question (Q1). For \textit{Background}, as students received feedback and gained more practice opportunities, the proportion of students who did \textit{Background} dimension correctly is significantly higher in Q3 as compared to Q1 (\textit{p} = .039). Additionally, we conducted a Pearson correlation between students' performance in the first practice task and their self-reported prior AI usage frequency. We found that students' frequency of using AI is positively correlated with their average score on the first question (\textit{r} = 0.27, \textit{p} = .017), indicating that students' prior access to AI might influence their initial prompt quality.

\begin{table*}[ht]
\fontsize{8.5pt}{10pt}\selectfont
\centering
\begin{tabular}{lcccccccc}
\hline
& & \textbf{Relevance} & \textbf{Purpose} & \textbf{Conciseness} & \textbf{Background} & \textbf{Elaboration} & \textbf{No Answer} & \textbf{Overall} \\ 
\hline
\textbf{Student Final Score (\textit{M})} & Q1 & 0.89 & 0.72 & 0.95 & 0.04 & - & - & - \\
(Applied human-graded & Q2 & 0.93 & 0.65 & 0.94 & 0.12 & 0.59 & - & - \\
scores, Yes-1/No-0) & Q3 & 0.89 & 0.75 & 0.95 & 0.11 & 0.52 & 0.64 & -\\
 & Overall & 0.90 & 0.71 & 0.95 & 0.09 & 0.56 & 0.64 & - \\
\hline
\end{tabular}
\caption{Student-written prompt scores by dimension. ``-" indicates this dimension is not applicable for that question.} 
\label{tab:stu_prompt_score}
\end{table*}

\subsubsection{Students are more confident in using AI to help learning after the learning activity.}
After the activity, students' self-reported confidence levels increased by 10.4\% on average (\textit{SD} = 0.92), and the Wilcoxon test result indicates the significance of such increment in students' confidence level (\textit{p} $<$ .001). Such improvement can also be observed from Figure \ref{fig:merged}, the significant rightward shift in the distribution of confidence scores indicates increased confidence in appropriate AI usage in learning.

\subsubsection{Lessons Learned: Students understood prompt basics, but their ability to create one varies.}
Ninety-eight students completed both the pre- and post-test. However, no significant increment (\textit{p} = .377) was found in the post-test (\textit{Mean} = 4.4, \textit{Median} = 5.0, \textit{SD} = 0.96, full score = 6.0) due to the ceiling effect in the pre-test (\textit{Mean} = 4.4, \textit{Median} = 5.0, \textit{SD} = 1.04). In other words, students had the basic skills to identify the prompt quality from a conceptual standpoint, although their prompt writing skills remain varied.

\subsection{RQ3: How do students perceive the benefits and challenges of this experience?}
We also examined student perceptions including what they learned from this activity, the interests and challenges encountered, and suggestions to refine this activity in the future. From survey responses from \textit{Study 1}, we found the following themes about students' learning experiences. 

\subsubsection{What do you learn from this learning activity?}
Most learners (87\%) reported that they learned AI-related knowledge, including \textit{\textbf{how to use AI in learning}} (e.g. \textit{“I can’t ask AI to write answers directly, otherwise I won’t learn anything."}), \textit{\textbf{how to ask AI questions effectively}} (e.g. \textit{“Adding background and context information in the prompt can help AI to form better answers"}), and \textit{\textbf{AI's capabilities}} (e.g. \textit{“AI can help on exam preparation", “AI also has problems that it doesn\'t know about"}). These indicated learners' gain on conceptual knowledge and skills of using AI in general and in learning, aligning with our instructional goals.

\subsubsection{What do you like about this learning activity?}
Fifty-five students showed great interests in \textbf{\textit{interaction with AI}}. More specifically, students appreciated the chances to ask questions, get responses and evaluate responses with newly learned prompting skills (e.g. \textit{“I am interested in the different responses from the different question-asking approaches"}). The \textbf{\textit{learning design}} was also favored by students. Many of them gained great interest in the learning objectives: \textit{“Knowing AI can be used to study or review coursework", “Great to know that AI can not only provide knowledge, but also create exercise questions."} Some students liked the in-time, comprehensive feedback design: \textit{“(the platform) will give detailed instructions after checking the answers I reply."} The scenario design was also favored by 8 students (e.g. \textit{“(I like) the scientific report of water cycle and want to know more about it"}). Additionally, some students favored the \textbf{\textit{visual elements}} of the learning platform, including the cartoon logos and its animations (e.g. \textit{“I like the cute cartoon robot holding a sign to tell me whether my answer is correct"}). In short, the parts that students favored highly overlap with our instructional goals.

\subsubsection{What is the most challenging part of this learning activity?}
Students' main challenges could be categorized into productive struggles \cite{warshauer2015productive} which are essential for learning, and elements that increase their extraneous load that should be minimized. The most common productive struggle was \textit{\textbf{experiencing difficulties in asking AI questions}}, which was the core skill we intended to teach in this activity. While students acknowledged their lack of competency in \textit{“writing a prompt that leads to a more helpful answer"}, many of their responses directly quoted some of the prompting guidelines in our instructional materials, which were strong indicators of students' mastery of tips for writing better prompts. For instance, students admitted the difficulties in writing prompts by saying \textit{“I didn't write a concise question", “I didn't provide context information"}. These precise quotes from the instruction content in survey responses indicated students' knowledge retention of prompting strategies, which served as evidence of our effective activity. Reasons for increasing extraneous load included \textbf{\textit{limitations of AI and learning platform}} (e.g. \textit{\textit{“It takes so long for AI to response", “I have trouble in logging in"}}), \textbf{\textit{lack of variation on scenario content}} (e.g. “All scenarios are in science subjects which I'm not interested in"), and noticeably, \textit{\textbf{limitations of students' basic computer skills}} (N = 22, e.g. \textit{“I am not good at typing, it is tedious to type a lot to ask AI questions"}). Some students struggled in typing, copying, and pasting information, which might hinder their prompt construction, even if they understood how to improve one.

\section{Study 2: Assessment Iteration Follow-Up}
Results from Study 1 revealed a ceiling effect in the pre-test, indicating that students in our study had a basic understanding to \textit{identify} learning-oriented prompts. However, practice performance showed that they had not yet mastered \textit{writing} such prompts effectively (Table \ref{tab:stu_prompt_score}). Therefore, to better measure their prompting literacy learning \cite{judson2012learning, staus2021addressing}, we iterated the assessment and conducted \textit{Study 2} to evaluate the assessment quality. Study 2 utilized the same procedure and materials with a similar population, except for the iterated assessment questions. Study 2 focused on RQ4 about assessment questions: 

\textbf{RQ4:} Does the revised assessment demonstrate improved quality in measuring students' prompting literacy?

\subsection{Iterated Assessment Design}
The iterated assessment included 15 questions: ten True or False (TF) questions and five Open-Ended (OE) questions. Specifically, TF1 to TF6 and OE1 targeted at LO-AI Capacity for Supporting Learning; TF7 to TF10 and OE2, OE3 targeted at LO-Contexts to Use AI for Learning; and OE4 and OE5 targeted at LO-Effective Prompt Formation.

\subsubsection{Replace multiple-choice questions with True / False and open-ended questions}
The lack of information about learners' prior knowledge can reduce the effectiveness of MCQ distractors. In contrast, TF questions require students to make a decision on each item individually, offering more data points to assess both the learners' true knowledge level and the quality of the question items, particularly at lower granularity \cite{brassil2019multiple}. In addition, to overcome difficulties in forming high-quality distractors in MCQs, one approach is to collect learners' common misconceptions as MCQ items. OE1, OE2, and OE3 are designed as a misconception collectors for \textit{LO-AI Capacity for Supporting Learning} and \textit{LO2-Contexts to Use AI for Learning} to facilitate further iterations on assessment questions. 

The gap between high MCQ scores and poor prompt-writing performance from Study 1 suggested that \textit{LO-Effective Prompt Formation} involves higher-order cognitive skills (falling under the \textit{analyze} and \textit{create} levels of Bloom's Revised Taxonomy \cite{krathwohl2002revision}), and developing MCQs that effectively assess these higher cognitive levels is challenging even for expert instructors in traditional disciplines \cite{douglas2012multiple}, let alone for an emerging topic. Therefore, we switched to open-ended questions and designed OE4 (\textit{Given a prompt and the AI Chatbot response, do you think the question can generate good learning material for your quiz preparation? Why or why not?}) to assess students' analyzing skill, and OE5 (\textit{Rewrite the question to AI to generate better preparation questions}) to assess prompt-writing skill more directly.

\subsubsection{Add abstract-level questions to increase the question difficulty variations} To more accurately estimate learners' prior knowledge, assessment questions should be designed with varying levels of difficulty and knowledge depth under the same learning objective. As such, abstract-level questions are included alongside concrete, scenario-based questions to assess learners across a broader spectrum of difficulty within the same objective.

\begin{figure*}[ht]
  \centering
  \includegraphics[width=\linewidth]{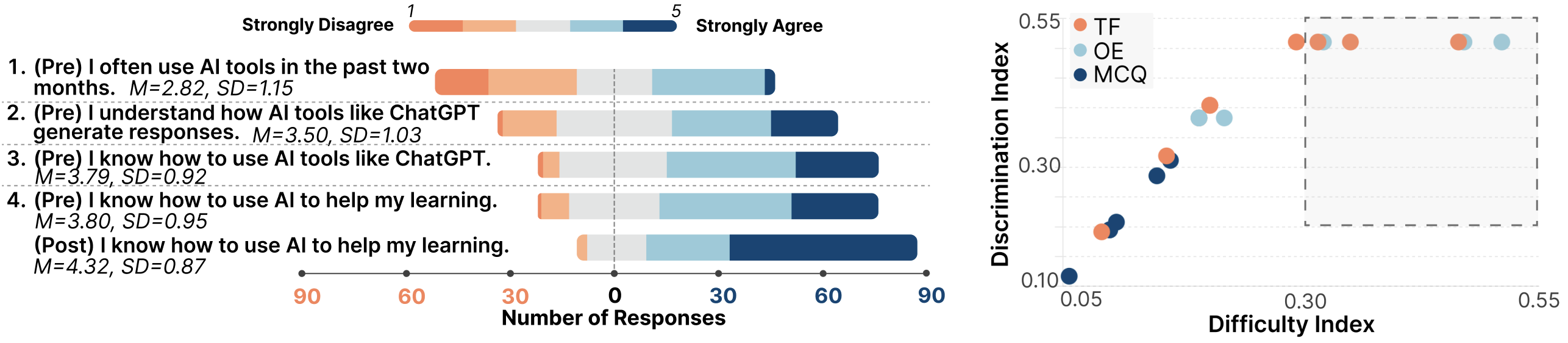}
  \caption{Left: Student responses to likert questions in the pre-test and post-test; Right: difficulty level vs discrimination index for all questions in both assessments. True or False (TF) questions and Open-Ended (OE) questions are in the revised assessment, and Multiple Choice Questions (MCQ) are in the original assessment.}
  \label{fig:merged}
\end{figure*}
\subsection{RQ4: Does the revised assessment better evaluate the targeted learning objectives?}

We evaluated the assessment question quality using difficulty and discrimination indices, and reported the overall reliability using Cronbach’s Alpha. The comparison of the initial and revised assessment is used to answer RQ4. In particular, we compared the quality of these two assessments using the data collected from two studies and three widely-used assessment evaluation matrices: difficulty, discrimination, and reliability \cite{nitko1996educational, salkind2017tests}. Our analyses yielded two preliminary findings:

\subsubsection{Item difficulty and discrimination are improved in the iterated version.}
The difficulty level and discrimination index are calculated based on the formula definitions provided in the previous work \cite{boopathiraj2013analysis}. Figure \ref{fig:merged} displays the difficulty level and discrimination index for all questions. A good discrimination index should be no less than 0.2 and ideally approach 1, while a good difficulty level falls within the range of [0.3, 0.7]. None of the MCQ questions fall into the desired range, while 60\% of OE and 30\% of TF questions satisfied the requirement, indicating an improvement on the item quality as assessment .

\subsubsection{Question reliability should be evaluated with a larger population later} With the initial success on difficulty level and discrimination index, Cronbach’s Alpha analysis showed moderate internal reliability for both assessment versions: 0.68 for the original and 0.58 for the iterated version. Although slightly below the 0.70 benchmark, these values are reasonable given the small sample size and limited number of items. The lower reliability in the iterated version may reflect greater item diversity or more targeted revisions. Larger-scale administrations will be needed to confirm these patterns and further strengthen internal consistency.

\section{Lessons Learned and Future Work}

In this work, we first designed and implemented a K-12 prompting literacy learning module featuring scenario-based chatbot interactions and AI-driven auto-assessment, and then conducted two classroom studies: \textit{Study 1} to measure technical capabilities and learning experience, and \textit{Study 2}, which built on data-informed refinements from \textit{Study 1} to evaluate the assessment quality. 

First, our results provided \textit{\textbf{design implications of teaching prompting literacy and AI}}. By experiencing the learning-by-doing prompting activities with LLM-generated immediate elaborated feedback in various scenarios, students gained more experience in including contexts in their prompts and increased confidence in prompt writing for learning. However, the lack of improvement in other aspects (e.g., conciseness, elaboration) and post-test performance may be attributed to dosage effects \cite{zhai2010dosage}, suggesting that more exposure is needed for aspects that students are less familiar with. Survey responses showed that task scenarios increased students’ interest and motivation, but students also wanted more relatable and diverse topics to match their varied interests. We also note that our study did not compare this approach to other AI literacy instructional methods; future work should conduct comparative studies or use this study as a benchmark for further iterations.

Secondly, this work introduced \textit{\textbf{a scalable learning and assessment platform}}. The AI-based auto-grader achieved high accuracy in most dimensions, indicating its potential to provide immediate, detailed feedback for learning prompt writing. However, certain limitations remain in the \textit{No Direct Answer} and \textit{Purpose} dimensions, suggesting future refinement. The AI's tendency to misinterpret the prompt intention  or overly focus on language mechanics highlights where human oversight and rubric refinement improve reliability.

Lastly, this work contributed to the development of learning and assessment materials for prompting literacy, informed by iterative design and data-driven refinement. As the first local study on prompting literacy, students’ baseline knowledge was unclear, making classroom-based iteration essential for ensuring instructional effectiveness. Increasing the variety and number of assessment items will help better surface learners’ misconceptions and knowledge gaps. These initial efforts establish a foundation for prompting literacy instruction. Future work should scale assessments to larger populations and items, using models such as Item Response Theory (IRT) \cite{embretson2000irt} to produce statistically robust measures of assessment quality.

\bibliography{aaai2026}

\end{document}